\documentclass[10pt,aps,pra,twocolumn,final,footinbib,amssymb,amsmath,amsfonts,showpacs,superscriptaddress,showpacs,floatfix]{revtex4-1} 
\usepackage{graphicx,times}

\graphicspath{{../plots_new/}}

\newcommand{\ket}[1]{|{#1}\rangle}
\newcommand{\bra}[1]{\langle{#1}|}
\newcommand{\aver}[1]{\langle{#1}\rangle}
\newcommand{\moy}[1]{\langle{#1}\rangle}
\newcommand{\elem}[3]{\langle{#1}\vert{#2}\vert{#3}\rangle}
\newcommand{\com}[2]{[{#1},{#2}]}
\newcommand{\bhat}{\hat{b}}
\def\Ham{\mathcal{H}}

\begin{document}

\title{Slow quench dynamics of Mott-insulating regions in a trapped Bose-gas}

\author{Jean-S\'ebastien Bernier}
\affiliation{Centre de Physique Th\'eorique, CNRS, \'Ecole Polytechnique, 91128 Palaiseau Cedex, France.}
\affiliation{Department of Physics and Astronomy, University of British Columbia, Canada V6T 1Z1.}
\author{Dario Poletti}
\affiliation{D\'epartement de Physique Th\'eorique, Universit\'e de Gen\`eve, CH-1211 Gen\`eve, Switzerland.}
\author{Peter Barmettler}
\affiliation{D\'epartement de Physique Th\'eorique, Universit\'e de Gen\`eve, CH-1211 Gen\`eve, Switzerland.}
\author{Guillaume Roux}
\affiliation{Laboratoire de Physique Th\'eorique et Mod\`eles Statistiques, Universit\'e Paris-Sud, CNRS, 
UMR8626, 91405 Orsay, France.}
\author{Corinna Kollath}
\affiliation{D\'epartement de Physique Th\'eorique, Universit\'e de Gen\`eve, CH-1211 Gen\`eve, Switzerland.}

\begin{abstract}
We investigate the dynamics of Mott-insulating regions of a trapped bosonic gas as the interaction strength 
is changed linearly with time. The bosonic gas considered is loaded into an optical lattice and 
confined to a parabolic trapping potential. Two situations are addressed: the formation of Mott domains in a 
superfluid gas as the interaction is increased, and their melting as the interaction strength is lowered. 
In the first case, depending on the local filling, Mott-insulating barriers can develop and hinder 
the density and energy transport throughout the system.  In the second case, the density and local energy 
adjust rapidly whereas long range correlations require longer time to settle. For both cases, we
consider the time evolution of various observables: the local density and energy, and their respective currents, 
the local compressibility, the local excess energy, the heat and single particle correlators. 
The evolution of these observables is obtained using the time-dependent 
density-matrix renormalization group technique and comparisons with time-evolutions done within 
the Gutzwiller approximation are provided.

\end{abstract}

\date{\today}
\pacs{05.70.Ln, 02.70.-c, 05.30.Rt, 67.85.Hj}

\maketitle

\section{Introduction}
Due to their good isolation from the environment and to their 
tunability, ultra-cold quantum gases are ideal candidates
to explore systems away from equilibrium~\cite{BlochZwerger2008}. 
Cold atoms are
well suited to explore situations where the 
Hamiltonian describing a system is slowly varied with time. 
Understanding the physical implications of such slow quenches is of great 
theoretical and practical importance to shed light on the coherent evolution 
of quantum systems and to devise methods to prepare complex quantum phases.
Seminal works on the dynamics of classical
systems near a second order phase transition conducted by
Kibble~\cite{Kibble1976} and Zurek~\cite{Zurek1985} identified that the defect production 
rate as a function of the ramp velocity is described by a scaling law when the system crosses a 
critical point. However, despite many recent theoretical 
advances~\cite{Dziarmaga2002, ClarkJaksch2004, Sengupta2004, ZurekZoller2005, Polkovnikov2005, 
Cucchietti2007, Canovi2009, Dziarmaga2010, Pollmann2010, Zimmer2010, Zimmer2011,
Trefzger2011, Polkovnikov2011, BernierKollath2011, Dora2011, Kennett2011, Delande2009,
DziarmagaBishop2002, SchuetzholdFischer2006, Fischer2006, Moessner2010, RodriguezSantos2009, DeGrandi2010, 
EcksteinKollar2009, MoeckelKehrein2010, EckardtHolthaus2005,EckardtLewenstein2011,PolettiKollath2011}, 
the response of strongly-correlated quantum gases to the slow 
quench of a Hamiltonian parameter is still far from being fully understood. Meanwhile, on the 
experimental side, considerable efforts have been devoted to understand the dynamics of 
interacting bosonic atoms when the depth of the optical lattice is 
varied~\cite{GreinerBloch2002, HungChin2010, Sherson2010, BakrGreiner2010, ChenDeMarco2011}
or when the slow quench of an effective parameter is 
performed~\cite{ZenesiniArimondo2009,StruckSengstock2011}. 

In relation to these experimental protocols, in Ref.~\onlinecite{HungChin2010, Rapp2010, NatuMueller2010, BernierKollath2011}, 
the presence of a parabolic trapping potential was found to significantly influence the dynamics. 
Two dynamical regimes have been shown to exist when interacting atoms loaded into an optical lattice and confined to a trap 
are subjected to a slow change of the interaction strength. For short ramp times, the evolution is dominated 
by intrinsic local dynamics, which is also present in a homogeneous system, whereas, for
longer ramp times the density redistribution can play an important role.

In this work, we study the response of bosonic atoms to a linear change of the interaction
strength. As these atoms are confined to one-dimensional tubes and loaded into an optical
lattice running along the tubes main axis, the physics for a wide range of parameters 
is well described by the one-dimensional
Bose-Hubbard model. Here our main objective is to understand the evolution, as a function of the
ramp time, of the local and non-local observables of the quantum gas. We focus on the crucial issues of the
formation and melting of Mott domains and on how the adiabatic limit is approached.

The article is structured as follows: In Sec.~\ref{sec:protocol}, we introduce the model and the time-dependent protocol. 
Sections~\ref{sec:methods} and \ref{sec:continuity} detail the methods and the theoretical definitions. These two sections can 
be skipped by readers more interested in the main phenomena. In Sec.~\ref{sec:SF_MI}, we turn 
to the description of the evolution resulting from the increase of the interaction strength. In Sec.~\ref{sec:two_dyn}, we focus on the 
occurrence of two dynamical regimes, the intrinsic dynamics and the dynamics induced by the 
trapping potential, and explain their origin (Sec.~\ref{sec:pert}). Afterwards, we direct our attention to the formation 
of ``Mott barriers'' which strongly block the equilibration process (Sec.~\ref{sec:form_MI}), the
energy transport (Sec.~\ref{sec:energy}) and the evolution of longer range correlations (Sec.~\ref{sec:corr}). 
Then, in Sec.~\ref{sec:MI_SF}, we consider the opposite case of melting the Mott domains occurring when the
interaction strength is lowered and, in particular, we point out the long equilibration times for long 
range correlation functions. For both situations, we characterize the time evolution considering various observables 
such as the density, the compressibility, the energy, various particle 
correlators, and the momentum distribution which is related to the interference patterns in time-of-flight 
measurements. These results are supplemented by a detailed 
analysis of the physical mechanisms responsible for the presence of the intrinsic and 
global dynamical regimes. We further show that the different time-scales can be identified experimentally
from interference patterns. Our numerical results are obtained from the time-dependent density-matrix
renormalization group method (t-DMRG). We also compare these quasi-exact results to time-evolutions done within 
the mean-field Gutzwiller method (Sec.~\ref{sec:gutz}) to identify the limitations of the
latter approach and pinpoint the qualitative physical insights it provides. The present 
work extends substantially our previous results on the same setup~\cite{BernierKollath2011}.
 
\section{Model and conservation laws}
\label{sec:setup}

\subsection{Hamiltonian and time-dependent protocol}
\label{sec:protocol}
Ultracold bosons in optical lattices are, in a wide parameter regime, well described by the Bose-Hubbard Hamiltonian~\cite{FisherGrinstein1988,JakschZoller1998}:
\begin{equation*}
\mathcal{H} = -J \sum_l \left(\bhat^{\dag}_{l+1} \bhat_l + \text{h.c.}\right)
        + \frac{U(t)}{2} \sum_l \hat{n}_l(\hat{n}_l-1) -\sum_l \mu_l \hat{n}_l\,,
\end{equation*}
with $\bhat^{\dag}_l$ the operator creating a boson at site $l$ and $\hat{n}_l=\bhat^{\dag}_l \bhat_l$ the local density operator. 
The total number of atoms is fixed to $N$.
The first term of the Hamiltonian corresponds to the kinetic energy of the atoms with the hopping amplitude $J$ and the second to the onsite interaction of strength $U$. 
The site-dependent chemical potential $\mu_l$ accounts for an external confinement. 
We consider here a one-dimensional geometry (tube) and use either a homogeneous ($\mu_l=0$) or a harmonic trapping potential of the form $\mu_l= -V_0(l-(L+1)/2)^2$, with $L$ the number of sites in the tube (open-boundary conditions) used in our simulations. We assume an experimentally realistic strength for the trapping potential of $V_0=0.006 J$ and particle number $N=24,48$. For these parameters the choice $L=64$ assures that edge effects are not important. 
One non-trivial aspect of the model is that it is non-integrable~\cite{Kolovsky2004, Kollath2010} for non-zero $J$ and $U$. Further, at commensurate fillings, a quantum phase transition from a superfluid to a Mott-insulating state occurs (at $(U/J)_c \approx 3.4$ for unity filling in one-dimension~\cite{Kuhner2000, ZakrzewskiDelande2008}). 
This phase transition is accompanied by the opening of a gap in the low-energy excitation spectrum, which strongly modifies the ground-state, thermodynamic and transport properties.
At incommensurate fillings a crossover between a superfluid and a Tonks-Girardeau, or hard-core boson, gas occurs in equilibrium. 
This distinct behavior at commensurate and incommensurate fillings implies that in a trapped system, different states can coexist in spatially separated regions~\cite{BatrouniTroyer2002, KollathZwerger2003, Folling2006}.
For instance, for strong enough interaction, a Mott-insulating plateau with commensurate filling, surrounded by a superfluid region, emerges.  

Regarding the time-dependent protocol, we consider a slow quench of the interaction strength $U(t)$ which can be achieved experimentally using a suitable Feshbach resonance~\cite{Inouye1998}. Different time-dependent protocols have been considered in previous works in homogeneous systems using several analytical or numerical approximation schemes~\cite{SchuetzholdFischer2006,DziarmagaBishop2002, Delande2009,Canovi2009,Cucchietti2007,RodriguezSantos2009,Dora2011,ClarkJaksch2004}.
For sake of simplicity and generality, the variation in time is chosen to be linear, starting from $U_i$ up to a final value $U_f$: $U(t) = U_i + \frac{t}{\tau}\delta U$ with $\tau$ the \emph{ramp time} and $\delta U = U_f-U_i$ the \emph{quench amplitude}. 
The real-time evolution starts from the ground state corresponding to $U_i$. 
The labels $i/f$ are used for the initial and final ground state values, respectively. 
The limit $\tau \rightarrow 0$, i.e. the sudden quench limit, has been studied intensively in the Bose-Hubbard model using analytical~\cite{AltmanAuerbach2002, Polkovnikov2002,  Fischer2006} and numerical methods~\cite{Kollath2007, LaeuchliKollath2008, Roux2009, Roux2010, Biroli2010, Sciolla2010, Sciolla2011}.

\subsection{Methods}
\label{sec:methods}
\subsubsection{t-DMRG}

Accurate ab-initio numerical simulations of the time evolution of the quantum gas are carried out using the t-DMRG technique~\cite{White1992, Vidal2004, WhiteFeiguin2004, DaleyVidal2004,Schollwoeck2005}. The time-evolution is implemented using the second order Trotter-Suzuki decomposition. The dimension of the effective space is a few hundred states and the time-step is adjusted with the ramp velocity. We introduce a cutoff value of $M=5$ or $6$ in the number of onsite bosons as higher boson occupancies are negligible in the situations considered here.

\subsubsection{Gutzwiller variational method}

In this section, we present how to determine the evolution of the system within the Gutzwiller 
mean-field method \cite{RokhsarKotliar1991,JakschZoller2002}. 
This approximation has been used before to describe the evolution during a slow change of the lattice depth in a higher dimensional trapped Bose-Hubbard system~\cite{Zakrzewski2005, NatuMueller2010}.
The Gutzwiller method is based on a variational ansatz of the many-body wave-function $|\Psi\rangle=\bigotimes_l \left[\sum_{n_l}c_{l,n_l}(t)|{n_l}\rangle\right]$  where $|{n_l}\rangle$ is the Fock state on site $l$ with $n_l$ particles and $c_{l,n_l}$ are the variational parameters.
The ground state for a given Hamiltonian is obtained by minimizing the total energy $E_{GW}$: 
\begin{eqnarray*}
E_\text{GW} &=& -J \sum_l \left( \moy{\bhat_l}\!^* \moy{\bhat_{l+1}} + \text{c.c.} \right) \\
&&+ \frac{U(t)}{2} \sum_{l,n_l} n_l (n_l-1) |c_{l,n_l}|^2  - \sum_{l,n_l} \mu_l n_l|c_{l,n_l}|^2,
\end{eqnarray*} 
where $\moy{\bhat_l} = \sum_{n_l}  \sqrt{n_l+1} \; c^*_{l,n_l}c_{l,n_l+1}$ and $^*$ denotes complex conjugation.  
The validity of the Gutzwiller method in evaluating static observables of one-dimensional systems has been studied, for example,
in Ref.~\onlinecite{Garcia-RipollvonDelft04}. 
The Gutzwiller approach predicts, in one-dimension, a phase transition at $(U/J)_c = 2~(1+\sqrt{2})^2 \simeq 11.7$~\cite{RokhsarKotliar1991}. 
The superfluid phase is signaled by a non-vanishing order parameter $\aver{\bhat_{l}}$ and, for a small interaction strength, the 
properties of local quantities are reasonably well approximated. 
In contrast, the Mott-insulating phase is characterized by a vanishing order parameter and vanishing local compressibility, 
thus neglecting completely particle fluctuations which are present in the real Mott-insulating phase. 
  
The time evolution for the coefficients $c_{l,n_l}(t)$ can be readily derived \cite{JakschZoller2002} from the Schr\"odinger equation. The equations are
\begin{eqnarray*}
i \hbar \partial_t c_{l,n_l}(t) &=& \left\{\frac{U(t)}{2} n_l (n_l-1) - \mu_l n_l\right\}c_{l,n_l}(t) \\ 
&& -J \sqrt{n_l+1} [\moy{\bhat_{l-1}}\!^* + \moy{\bhat_{l+1}}\!^*]\,c_{l,n_l+1}(t) \\
&& -J  \sqrt{n_l} [\moy{\bhat_{l-1}} + \moy{\bhat_{l+1}}]\,c_{l,n_l-1}(t) \,.   
\end{eqnarray*} 
They are solved numerically by implementing a split-step method.    

\subsection{Observables and definitions from continuity equations}
\label{sec:continuity}
\subsubsection{Correlations and interference pattern}

We define the one-body correlation function between sites $l$ and $m$
as
\begin{equation}
\label{eq:greens}
g_{l,m}= \frac{1}{2}\aver{\bhat_l^\dagger\bhat_{m} + {\rm h.c.}} \;.
\end{equation}
While $g$ is not easily accessible, time-of-flight techniques measure
an interference pattern related to the momentum distribution of the
correlated gas.  Neglecting the Wannier-function envelope, the
interference pattern is given by
\begin{equation}
N(k) = \frac 1 L \sum_{l,m} e^{i(l-m)ka} \moy{\bhat_l^\dag \bhat_m}
\label{eq:nk}
\end{equation}
where $a$ is the lattice spacing.
For a superfluid state, $N(k)$ is expected to be strongly peaked
around zero momentum, whereas for a Mott-insulating state the
interference pattern should be rather flat.

\subsubsection{General expression for the continuity equation}

In this section, we want to determine, in the Schr\"odinger picture, the current operators corresponding to a given observable $O(t)=\elem{\psi(t)}{\hat{O}(t)}{\psi(t)}$, where $\hat{O}(t)$ can explicitly depend on time, using the associated continuity equation. 
In the following, the shorthand notation $\moy{\cdots}$ stands for the expectation value $\elem{\psi(t)}{\cdots}{\psi(t)}$ at a given time.
The continuity equation takes, on general grounds, the following form: 
\begin{align}
\hbar \partial_t O(t) \label{eq:continuity}
= -\text{div}\moy{\hat{J}^{O}} + \moy{\hat{S}^O}\,.
\end{align}
$\hat{J}^{O}$ is the current operator for which we have
\begin{subequations}
\begin{align}
-\text{div} \moy{\hat{J}^{O}} &=i \moy{\com{\Ham(t)}{\hat{O}(t)}}\\ 
&= -(\moy{\hat{J}^{O}_{l,l+1}} - \moy{\hat{J}^{O}_{l-1,l}}) \,. \label{eq:1dspez}
\end{align}
\end{subequations}
The second equality is the specialization to a one-dimensional lattice for an observable located around site $l$, with incoming and outgoing currents. 
As we are interested in studying a one-dimensional system, \eqref{eq:1dspez} is used throughout. 
The source operator $\hat{S}^O(t)=\hbar\partial_t\hat{O}(t)$ is non-zero only for an explicitly time-dependent operator.

Interestingly, integrating \eqref{eq:continuity} between times $0$ and $\tau$, taking the adiabatic limit $\tau\rightarrow\infty$ and 
doing the change of variable $t\rightarrow U$ (in the source term integral) enables one to express the integrated contribution of currents only as a function of ground state expectation values
\begin{equation}
\int\limits_0^{\infty} \frac{dt}{\hbar} \,\text{div}\moy{\hat{J}^{O}}(t) = O_i - O_f + \int\limits_{U_i}^{U_f}\!\!dU\,\elem{\psi_0(U)}{\partial_U\hat{O}}{\psi_0(U)}\;,
\label{eq:adiabatic-O}
\end{equation}
where $\ket{\psi_0(U)}$ is the ground state corresponding to $U$. 
The integral on the right-hand side is taken along the adiabatic path.
This remark is important from a numerical perspective because the right-hand side can be efficiently 
computed via ground state techniques while the left-hand side would require time-dependent simulations 
over very long times, which is not feasible. 

As explained in the introduction, the main objective of this work is to characterize how particles and energy redistribute when the interaction strength, $U(t)$, is ramped up or down. Therefore, we introduce below the relevant quantities and the physically significant terms of their associated 
continuity equations. A few commutators, useful in the derivation of these continuity equations, are provided in Appendix~\ref{app:commutators}.

\subsubsection{Local observables on sites and bonds}

As a first example, \eqref{eq:continuity} can be used to derive the particle current associated with the local density $\hat{n}_l$:
\begin{equation}
\hat{j}_{l,k}\equiv \hat{J}^{n_{l}}_{l,k} = iJ(\bhat^\dagger_{k}\bhat_{l}-\bhat^\dagger_{l}\bhat_{k}).
\label{eq:partcurrent}
\end{equation}
This current is defined between sites $l$ and $k$ and there is no source term associated with $\hat{n}_l$ as it is 
not explicitly time-dependent. 
As the particle current appears quite often in the rest of this article, from now on, it will be 
denoted as $\hat{j}_{l,k}$. Finally, it is instructive to note that for a homogeneous and translationally invariant system, 
the local density is constant at all time due to the conservation of the total number of particles. 

In order to better understand the different time-scales involved during the evolution, it is also useful to consider separately 
the evolution equation for the density fluctuations $\hat{n}_l^2$.
This quantity is essential to our comprehension of the Bose-Hubbard model and is related to the the local compressibility $\kappa_l = \aver{\hat{n}_l^2}-\aver{\hat{n}_l}^2$. 
Using \eqref{eq:continuity} and after some algebra, we find that the evolution of $\hat{n}_l^2$ is controlled by a ``density-assisted'' or ``correlated'' current
\begin{equation}
\label{eq:evol_nn}
\hat{J}^{n^2_l}_{l,l+1}= \hat{n}_l\hat{j}_{l,l+1} + \hat{j}_{l,l+1}\hat{n}_l \; 
\end{equation}
(note that $\hat{J}^{n^2_l}_{l-1,l} =\hat{n}_l\hat{j}_{l-1,l} + \hat{j}_{l-1,l}\hat{n}_l$).
The origin of this density-assisted current, mixing $\hat{j}$ and $\hat{n}^p$ operators, comes from the 
evolution equation for the onsite occupancy probabilities discussed in Appendix~\ref{app:proba}. 
In equilibrium, in our system, the average of~\eqref{eq:evol_nn} computed in 
the ground state vanishes, as for the particle current operator. 

The same strategy is used to get the current operators associated with observables living on bonds, such as the local kinetic energy operator $\hat{K}_{l,l+1}$ (or nearest-neighbor one-particle correlation) defined by
\begin{equation}
\hat{K}_{l,k} = -J(\bhat_k^{\dagger}\bhat_l + \bhat_l^{\dagger}\bhat_k)
\label{eq:kinetic}
\end{equation}
between site $l$ and $k$.
We find that the incoming current associated with $\hat{K}_{l,l+1}$ reads
\begin{subequations}
\label{eq:fullkineticcurrent}
\begin{align}
\hat{J}_{l-1,l}^{K_{l,l+1}} =& -\mu_{l} \hat{j}_{l,l+1} \label{eq:kin_mu} \\
               & + J \hat{j}_{l-1,l+1}   \label{eq:kin_corr2} \\ 
               & + \frac{U(t)}{2} (\hat{n}_l\hat{j}_{l,l+1}+\hat{j}_{l,l+1}\hat{n}_l)\;, \label{eq:kin_densassisted}
\end{align}
\end{subequations}
showing the interplay of the correlated and usual particle currents.
It is worth noticing that for a homogeneous system, the evolution of local kinetic fluctuations is directly related to that of the density fluctuations since in this case 
\begin{equation}
\label{eq:n2-kin-relation}
\partial_t \aver{\bhat_l^\dagger\bhat_{l+1}+\bhat_{l+1}^\dagger\bhat_l}= \frac{U(t)}{2J} \partial_t \aver{\hat{n}_l^2}\;.
\end{equation}
Thus, even in the homogeneous limit, the time dependence of $U(t)$ affects the evolution of the local kinetic term or nearest-neighbor correlations. 
Eq.~\eqref{eq:n2-kin-relation} is also straightforwardly obtained from the evolution of the total energy discussed below. 
Note, this equation is not valid for inhomogeneous gases where the balance of particle currents can be non-zero. 

Similarly, the current operator associated with the particle current itself contains density-assisted hoppings, following the expression
\begin{subequations}
\label{eq:fullcurrentcurrent}
\begin{align}
\hat{J}_{l-1,l}^{j_{l,l+1}} =& \mu_{l} \hat{K}_{l,l+1} \label{eq:contr_mu_curr} \\
               & + 2 J^2 \hat{n}_l       \label{eq:contr_n_curr} \\
               & + J \hat{K}_{l-1,l+1}   \label{eq:contr_corr2_curr} \\ 
               & - \frac{U(t)}{2} (\hat{n}_l\hat{K}_{l,l+1}+\hat{K}_{l,l+1}\hat{n}_l)\;. \label{eq:contr_densassisted_curr}
\end{align}
\end{subequations}
We also give for clarity the outgoing current operator: $
\hat{J}_{l+1,l+2}^{j_{l,l+1}} $ $
=\mu_{l+1}\hat{K}_{l,l+1}+2J^2\hat{n}_{l+1}+J\hat{K}_{l,l+2} $ $
-\frac{U(t)}{2}(\hat{n}_{l+1}\hat{K}_{l,l+1} $ $
+\hat{K}_{l,l+1}\hat{n}_{l+1}) $. It is worth mentioning that the
correlated current and hopping terms in \eqref{eq:fullkineticcurrent} and \eqref{eq:fullcurrentcurrent}
all come with the interaction strength as a prefactor and disappear
for a non-interacting gas. Their behavior is thus strongly affected by the presence of
interactions. Finally, as in the next section the time-derivative of the particle current 
will be of great use to understand the mechanisms responsible for the evolution of the 
density profile, we provide here its full expression:
\begin{subequations}
\label{eq:diffcurrent}
\begin{align}
\hbar \partial_t \moy{\hat{j}_{l,l+1}} =& (\mu_{l}-\mu_{l+1}) \moy{\hat{K}_{l,l+1}} \label{eq:contr_mu} \\ 
& + 2 J^2 (\moy{\hat{n}_l} - \moy{\hat{n}_{l+1}}) \label{eq:contr_n} \\
& + J (\moy{\hat{K}_{l-1,l+1}} - \moy{\hat{K}_{l,l+2}}) \label{eq:contr_corr2} \\
& -\frac{U(t)}{2} \moy{(\hat{n}_{l}-\hat{n}_{l+1})\hat{K}_{l,l+1} + \text{h.c.}}\;.\label{eq:contr_densassisted}
\end{align}
\end{subequations}

\subsubsection{Energy and heat}
\label{sec:energy}
We now turn to the transport of energy by first defining the bond-symmetric local energy operator as
\begin{equation}
\label{eq:local_e}
\hat{h}_l= \frac{1}{2}[\hat{K}_{l-1,l} + \hat{K}_{l,l+1}] + U(t)\hat{I}_l -\mu_l \hat{n}_l \; ,
\end{equation}
where $\hat{I}_l=\hat{n}_l(\hat{n}_l-1)/2$ is the operator related to the interaction energy. 
In this case, we find that the energy current $\hat{J}^{h_l}_{l-1,l}$ is given by 
\begin{subequations}
\label{eq:fullenergycurrent}
\label{eq:allenergy}
\begin{align}
\hat{J}^{h_{l}}_{l-1,l} =& -\frac{\left(\mu_{l-1}+\mu_l\right)}{2} \hat{j}_{l-1,l}  \label{eq:energy_mu}\\ 
& -\frac{U(t)}{2} \hat{j}_{l-1,l}                                             \label{eq:energy_j}\\ 
& -\frac{J}{2} (\hat{j}_{l-2,l}  + \hat{j}_{l-1,l+1})                         \label{eq:energy_corr2}\\ 
& +\frac{U(t)}{4} \left[ (\hat{n}_{l-1} + \hat{n}_l)\hat{j}_{l-1,l} + \hat{j}_{l-1,l}(\hat{n}_{l-1}+\hat{n}_l) \right]\,,\label{eq:energy_densass}
\end{align}
\end{subequations}
in which we naturally recover the particle and correlated currents appearing in \eqref{eq:partcurrent}, \eqref{eq:evol_nn} and \eqref{eq:fullkineticcurrent}.
In addition, since the energy operator is explicitly time-dependent and therefore not a conserved quantity during the protocol, we have the following source term
\begin{equation}
\label{eq:sourceE}
\hat{S}_l^{h_l} = \hbar \partial_t U(t) \hat{I}_l
\end{equation}
which shows the importance of the density fluctuations in the energy production. 
In particular, the total energy $E(t)= \moy{\mathcal{H}(t)}$ satisfies the relation
\begin{equation}
\partial_t E(t) =\elem{\psi(t)}{\partial_t\Ham}{\psi(t)} = \left[ \partial_t U(t) \right] \sum_l \moy{\hat{I}_l}(t)\;,
\label{eq:total-energy}
\end{equation}
i.e., the energy put in the system is directly related to the evolution of the density fluctuations. 
For an inhomogeneous system, there are two contributions to the local energy production as seen 
from \eqref{eq:allenergy} and \eqref{eq:sourceE}: one from currents and correlated currents and one from the external driving of the system.
Summing up the total energy, the contribution from currents must vanish to fulfill \eqref{eq:total-energy}, but locally, one may have energy redistribution.
We can define the heat produced in the system as the energy of the atoms at the final time compared to that of the ground state for the final interaction strength  
\begin{eqnarray}
Q(\tau) &=& E(\tau)-E_{0,f} \nonumber \\ 
&=& E_{0,i}-E_{0,f} + \frac{\delta{U}}{\tau} \int_{0}^{\tau} dt \sum_l \moy{\hat{I}_l}(t)\;,
\label{eq:heat}
\end{eqnarray}
with $E_{0,i/f}$ the ground state energies.  
Note that $ \moy{\hat{I}_l} $ is accessible experimentally, which
makes it possible to measure the interesting $Q(\tau)$ dependence.
 We can quickly check that this formula gives back the
correct results in the sudden quench and adiabatic quench limits. 
In the sudden quench limit, $\ket{\psi(t)} = \ket{\psi_0(U_i)}$ which yields
$Q(0) = E_{0,i}-E_{0,f} + \delta U \sum_l \moy{\hat{I}_l}_{0,i}$.
This means that the heat only depends on ground state properties of the corresponding initial and final parameters. 
In the adiabatic case, we have
$\ket{\psi(t)}=\ket{\psi_0(U(t))}$ along the adiabatic path so the
integral can be reexpressed as $\int_{U_i}^{U_f} dU \sum_l
\moy{\hat{I}_l}_0(U)$, with $\moy{\hat{I}_l}_0(U) = \elem{\psi_0(U)}{\hat{I}_l}{\psi_0(U)}$. 
Using Feynman-Hellmann theorem over $U$, it is
clear that this integral cancels $E_{0,i}-E_{0,f}$ to make
$Q(\infty)=0$.

One can define a local excess energy $q_l$ as the difference in local
energies between the final energies and the ground state expectation
for the final parameters:
\begin{equation}
\label{eq:heatdef}
q_l(\tau) = \moy{\hat{h}_l}(\tau) - \moy{\hat{h}_l}_{0,f} \;.
\end{equation}
The local excess energy produced splits up into three different contributions
\begin{subequations}
\label{eq:allheat}
\begin{align}
\label{eq:heatdiff}
q_l(\tau) =& \moy{\hat{h}_l}_{0,i} - \moy{\hat{h}_l}_{0,f}\\
\label{eq:heatdiv}
& -\frac{1}{\hbar}\int_{0}^{\tau}dt \;\text{div}{\moy{\hat{J}^h}(t)} \\
\label{eq:heatfluct}
& + \frac{\delta U}{\tau}\int_{0}^{\tau} dt\; \moy{\hat{I}_l}(t)\;,
\end{align}
\end{subequations}
where the first term is simply the local ground state energies difference (independent of $\tau$), 
the second term is the integrated contribution of energy currents, and the last term is the integrated 
contribution due to the external operator. While $Q(\tau)$ is necessarily non-negative, $q_l(\tau)$ can 
be negative or positive depending on the relative contributions of each term.

Finally, using \eqref{eq:adiabatic-O} with $\hat{O} = \hat{h}_l$ allows one to 
calculate these quantities in the adiabatic limit:
\begin{equation}
\int\limits_0^{\infty} \frac{dt}{\hbar} \,\text{div}\moy{\hat{J}^{h_l}}(t) 
= \moy{\hat{h}_l}_{0,i} - \moy{\hat{h}_l}_{0,f} + \int\limits_{U_i}^{U_f}\!\!dU\,\moy{\hat{I}_l}_0(U)\;,
\label{eq:adiabatic-energy}
\end{equation}
where the right-hand side can be computed accurately using numerical techniques.

With this set of equations in mind, we are now ready to identify the different driving 
forces responsible for the system evolution when the interaction strength 
is ramped up or down.

\section{Digging a Mott domain in a superfluid}
\label{sec:SF_MI}

\subsection{Evolution of local quantities from t-DMRG}

In this section, we consider a linear quench from $U_i=4J$ to $U_f=6J$. 
$U_i$ is close to the homogeneous superfluid-Mott transition point and $U_f$ lies deeper in the Mott-insulating regime. 
We compare two typical situations: (i) the number of particles is chosen low enough in order for the maximal 
filling to remain below unity at all times ($N=24$); (ii) $N$ is sufficiently large so that, at $U_f$, the corresponding 
ground state density profile has a Mott-insulating ``shell'' and a superfluid center ($N=48$). 
We focus on different aspects of the dynamics: time-scales, role of insulating domains on particle 
transport, energy production and transport, and their experimental signature.

\begin{figure}[t]
\centering
\includegraphics[width=\linewidth,clip=true]{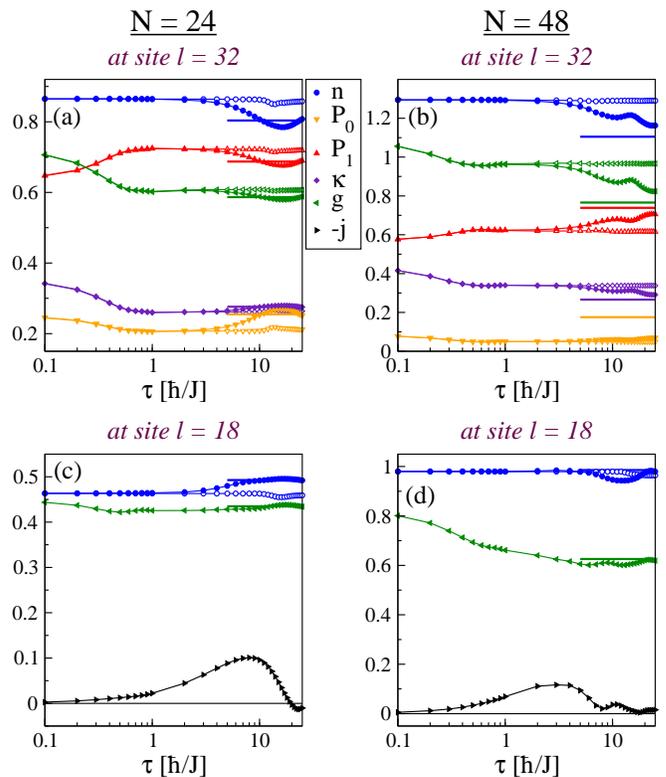}
\caption{(color online). Slow quench from $U_i = 4J$ to $U_f = 6J$.
Evolution of local observables in the presence of a trap as a function of the ramp time 
$\tau$, and compared with that of a homogeneous system (open symbols) having the same initial local density.
Observables are the density $n_l$, compressibility $\kappa_l$, occupancy probabilities $P_0$ and 
$P_1$, neighboring correlation $g_{l,l+1}$ and the particle current $j_{l,l+1} = \moy{\hat{j}_{l,l+1}}$.
Subplots correspond to two different total number of particles $N=24$ and $48$, and two 
different sites: $l=18$ and the central site $l= 32$ (cf.~Fig.~\ref{fig:trap_profiles_4_6} for the location of these sites).
\label{fig:local_vs_tau_4_6}}
\end{figure}

\subsubsection{Existence of two dynamical regimes}
\label{sec:two_dyn}
In Fig.~\ref{fig:local_vs_tau_4_6}, the final values ($t=\tau$) of most of the local observables introduced before 
(density, local compressibility, local particle current, local correlation $g_{l,l+1}$) and also the first two 
occupancy probabilities $P_0$ and $P_1$, are presented as a function of the ramp time $\tau$.
This figure clearly uncovers the existence of two dynamical behaviors.
First, we observe that for short ramp times, the densities at $l=32$ which lies in the center of the trap and $l=18$ which lies close to the forming Mott-insulating barrier are both approximately constant, following the evolution of the homogeneous system~\cite{Note1}.
In fact, variations (and oscillations) of both central and outer densities become significant only for longer ramp times, beyond $\hbar/J$. 
In contrast, the evolution of both the occupancy probabilities and the compressibility occurs on a much faster time-scale: these observables vary rapidly at short $\tau$ and display less pronounced variations at larger $\tau$.
These two distinct behaviors reveal the presence of two dynamical regimes~\cite{NatuMueller2010, Rapp2010, BernierKollath2011}:
(i) the intrinsic dynamics, here occurring at short-times before the particle transport sets in (present in both the homogeneous and trapped systems);
(ii) a long-time behavior associated with particle transport and clearly due to the inhomogeneous structure of the density profiles.

Qualitatively, one can understand the origin of different time-scales from the continuity equations of Sec.~\ref{sec:continuity}.
For instance, the incoming and outgoing particle currents balance each other in a translational invariant configuration so that the density remains constant.
However, for the density fluctuations whose current operator~\eqref{eq:evol_nn} has correlated terms, no such balance 
is achieved and consequently these quantities evolve with time.
In the case of an inhomogeneous gas, gradients of local quantities and chemical potentials inevitably give rise to particle currents, themselves sustaining the evolution of all local quantities.
Contrary to the intrinsic dynamics, we expect these effects to vanish with reducing the trap amplitude $V_0$. 
Thus, their time-scale is distinct from the intrinsic one and is related to the external potential strength.
In order to quantify better these ideas, we now present arguments based on perturbative calculations.

\subsubsection{Insights from perturbative expansions}
\label{sec:pert}
In Ref.~\onlinecite{BernierKollath2011}, we observed that, for the quench
parameters typically considered, the
homogeneous dynamics of most local observables was well reproduced by
time-dependent perturbation theory, particularly in the small-$\tau$ regime. Working in the
\emph{initial Hamiltonian} eigenstates basis $\ket{\alpha}$, of energy
$E_{\alpha}$, the first-order expansion in $\delta U / \tau$ for a real,
symmetric and dimensionless observable $\hat{O}$ reads:
\begin{equation}
\label{eq:pert-O}
O(\tau,\delta U) = O_{00} - 2 \frac{\delta U}{\hbar\tau} \sum_{\alpha \neq 0} \frac{\omega_{\alpha}\tau-\sin(\omega_{\alpha}\tau)}{\omega_{\alpha}^2} O_{0\alpha}I_{\alpha 0}\;.
\end{equation}
The frequencies $\omega_{\alpha} = (E_{\alpha}-E_0)/\hbar$ are excitation
energies of level $\ket{\alpha}$ with respect to the ground-state $\ket{0}$.
$I_{\alpha\beta}=\sum_l\elem{\alpha}{\hat{I}_l}{\beta}$ are the matrix
elements of the interaction operator and
$O_{\alpha\beta}=\elem{\alpha}{\hat{O}}{\beta}$ those of the
observable. Such series can be well-behaved in the thermodynamic
limit even in the absence of a spectral gap, and this is what we
observe for our setup by looking at different system sizes.  Taking
the $\tau \rightarrow 0$ limit~\cite{Note2},
the response of
the observable is typically quadratic
\begin{equation}
O(\tau,\delta U) \simeq O_{00}\left(1 \pm \frac{1}{2}f_{O}\tau^2\right)\,,
\end{equation}
where the $\pm$ sign depends on the observable. We have here introduce
the ``curvature''
\begin{equation}
f_O= \frac{2}{3}\frac{\delta U}{J} \frac{\tau_O^{-2}}{O_{00}}
\label{eq:fO}
\end{equation}
containing an intrinsic characteristic ramp time $\tau_O$
associated with the observable $O$ ($J$ is there for dimensionality
normalization):
\begin{equation}
\tau_O^{-2} = \frac{J}{\hbar} \Big\vert \sum_{\alpha \neq 0} \omega_{\alpha} O_{0\alpha}I_{\alpha 0}\Big\vert\;.
\label{eq:tauO}
\end{equation}
The curvature helps understand the departure from the initial value
$O_{00}$, as one can see, for example, in
Fig.~\ref{fig:local_vs_tau_4_6}. In particular, $f_O$ is linear with the
quench amplitude $\delta U$ (within this approximation).  
For example, in the homogeneous gas limit, one can obtain an explicit expression
for the driving of the particle fluctuations $f_{n^2}$ by, in addition, resorting to 
perturbation theory in $J/U_i$ (strong interaction limit).
We find that
\begin{equation}
f_{n^2}\approx \frac{32}{3}\frac{\delta U J^2}{\hbar^2 U_i}\;, 
\label{eq:fn2}
\end{equation}
which is consistent with the change of the compressibility at short
time plotted in Fig.~\ref{fig:local_vs_tau_4_6}. The breakdown 
of the quadratic behavior, which coincides with the onset of the relaxation, 
is expected to happen on a time scale $\tau=|f_{n^2}|^{-1/2}$. We note that the
parameter $J/U_i$ in Fig.~\ref{fig:local_vs_tau_4_6} is not in the
regime where perturbation theory is expected to give a quantitative
description. Nevertheless, putting numerical values in
\eqref{eq:fn2}, one finds short relaxation times (below $\hbar/J$)
compatible with Fig.~\ref{fig:local_vs_tau_4_6}.

It is also worth mentioning that in the definition of the curvature
\eqref{eq:fO}, we were careful to separate what depends 
on the quench protocol, the
parameter $\delta U$ and the prefactor $2/3$, from what is intrinsic
to the initial ground-state: $O_{00}$ and $\tau_O$. 
Indeed, when the $U(t)$ function is of the general
type $\delta U f(t/\tau)$, the prefactor $2/3$ is replaced by $4~\int_0^1
dx(1-x)f(x)$. Hence, $\tau_O$ is
an intrinsic characteristic time of the initial state. We stress that the
quantities in \eqref{eq:fO} and \eqref{eq:tauO} are accessible by
ground state numerical techniques. Within this perturbative framework,
one can easily understand the two-regimes discussed above and also 
derive, in the homogeneous case, relations between the characteristic 
time-scales of various observables.

We first consider the characteristic time associated with the local density 
operator. In the homogeneous case, the ground state is characterized by a
spatially uniform local density. Taking advantage of this symmetry, we
find that $\tau_n = \infty$. This result agrees with the fact that the 
density remains constant for all times.
In contrast, for the local density fluctuations (or
compressibility) and local kinetic energy, $\tau_{n^2/g}$ are finite
even in a homogeneous system since the matrix elements in
\eqref{eq:tauO} do not vanish. Furthermore, the two time-scales are
actually related to each other. Since the Hamiltonian has only two
terms, we find that $\sum_l \elem{0}{\hat{g}_{l,l+1}}{\alpha} = (
U_i/J)\sum_l\elem{0}{\hat{n}_l^2}{\alpha}$ leading to $\tau_g =
\sqrt{\frac{J}{U_i}} \tau_{n^2}$. 
This relation agrees with a dimensional
analysis of \eqref{eq:n2-kin-relation} and is also in qualitative
agreement with Fig.~\ref{fig:local_vs_tau_4_6}, where we find a
slightly slower relaxation for the kinetic term as compared to the
compressibility. In addition, these time-scales are themselves
related to the characteristic ramp time for the heat, $\tau_c$,
defined as $\tau_c^{-2}
= \frac{J}{12\hbar L}\sum_{\alpha} \omega_{\alpha}
\left\vert{I_{\alpha0}}\right\vert^2$~\cite{BernierKollath2011}.  
Then, we find that $\tau_{n^2}
= \tau_c/\sqrt{24}$ (although the prefactor depends on the chosen
definition for $\tau_c$).

We now turn to the situation where a trapping confinement is present. In this case,
the translational symmetry is lost leading to a finite $\tau_n$.
Naturally, $\tau_{n^2/g}$ should also be affected by the presence of the trap,
but provided the latter is small enough, the corrections can be
negligible as illustrated by Fig.~\ref{fig:local_vs_tau_4_6}.
We expect that $\tau_n(V_0)$ diverges when the trap magnitude $V_0$
reaches zero. Consequently, by tuning $V_0$ to a low enough value, one
should in general be able to observe the intrinsic dynamics of the
system occurring below $\tau_n$. The behavior of the $\tau_n(V_0)$
function is an open issue, particularly because $V_0$ is not a
perturbation in experiments and realistic numerical calculations.  If
we were to trust a naive first order perturbation argument for the
relatively unphysical situation of a gas in a box and perturbed by a
small $V_0$, one would expect a linear scaling of the matrix elements,
yielding the scaling $\tau_n \propto 1/\sqrt{V_0}$. 
However, this scaling only serves as an illustration of the above statements. 
Of course, $\tau_n$ also depends on $U_i/J$. Finally, we may argue that when 
$V_0$ is too large, transport phenomena could eventually hide the intrinsic dynamics.

These results put on firmer grounds the existence of two
different dynamical regimes: one deeply connected to
inhomogeneities and controlled by $V_0$, and the intrinsic one
present in the homogeneous gas and much less sensitive to $V_0$.

\subsubsection{Profiles and ``Mott barriers''}
\label{sec:form_MI}
\begin{figure}[t]
\centering
\includegraphics[width=\linewidth,clip=true]{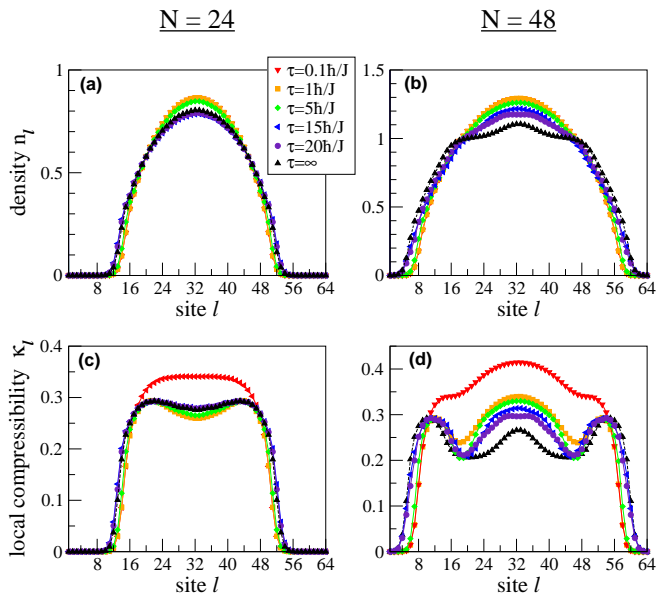}
\caption{(color online). Final local density and compressibility profiles after a slow quench 
from $U_i=4J$ to $U_f=6J$ for different ramp times, $\tau$, and for the ground state ($\tau = \infty$) 
at $U=6J$ in the trapped system. Left panels $N=24$, right panels $N=48$.
\label{fig:trap_profiles_4_6}}
\end{figure}
\begin{figure}[t]
\centering
\includegraphics[width=0.75\linewidth,clip=true]{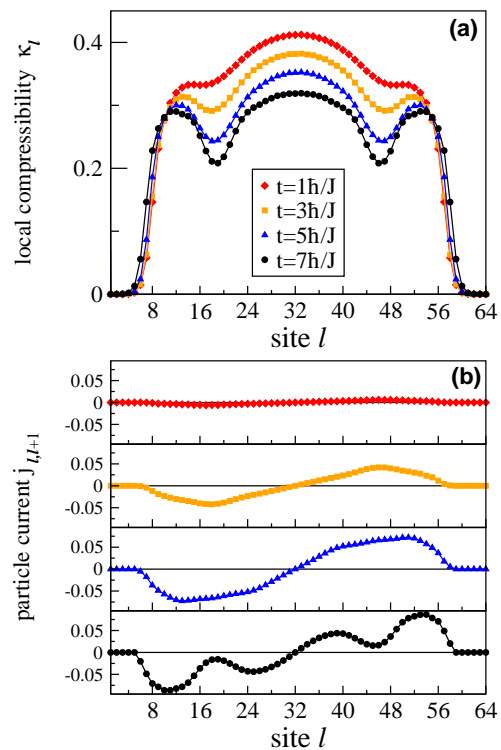}
\caption{(color online). Time-evolution of the local compressibility $\kappa_l$ and current $j_{l,l+1} = \moy{\hat{j}_{l,l+1}}$
during a slow quench from $U_i=4J$ to $U_f=6J$ in a time $\tau = 7\hbar/J$ for $N=48$ in the trapped system. 
As the ``Mott barriers'' are formed, the current in their vicinity weakens.
\label{fig:comp_vs_time}}
\end{figure}

We detail here the spatial evolution of local quantities.
In Fig.~\ref{fig:trap_profiles_4_6}, we present the final profiles for the density and compressibility as a function of ramp time. 
At low filling ($N=24$), the shape of these final profiles is well understood if one resorts to the arguments presented above: 
we see that for short ramp times the density profiles barely evolve while the compressibility changes considerably. 
For longer $\tau$, the profiles approach smoothly the final ground state configuration.
For the larger filling $N=48$, the evolution is more involved. 
A strong reduction of the compressibility occurs locally in regions of 
filling close to unity already for short ramp times $\tau \approx \hbar/J$ while
the formation of pronounced Mott-insulating ``shells'' in the density profile only takes place at much longer ramp times, after $5\hbar/J$.

\begin{figure}[t]
\centering
\includegraphics[width=\columnwidth,clip=true]{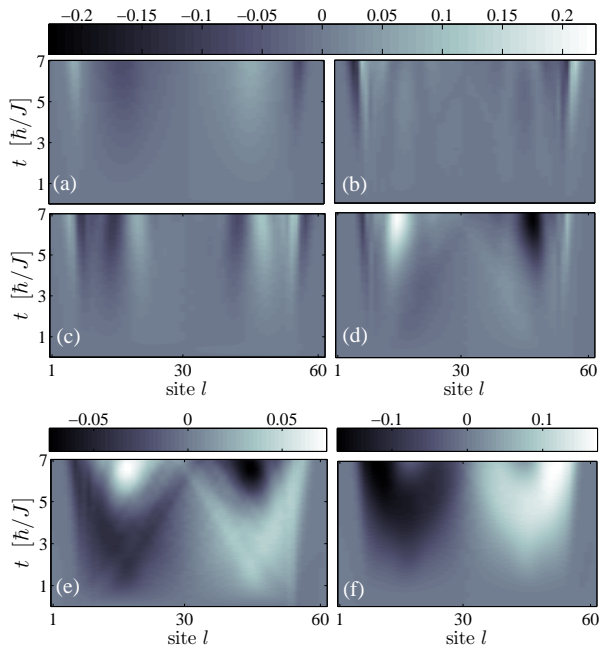}
  \caption{$(a-d)$ Time-evolution of the different contributions to the time-derivative of the particle current: 
Eq.~\eqref{eq:contr_mu} $(a)$, Eq.~\eqref{eq:contr_n} $(b)$, Eq.~\eqref{eq:contr_corr2} $(c)$ 
and Eq.~\eqref{eq:contr_densassisted} $(d)$. $(e)$ time-evolution of the time-derivative of the particle 
current. $(f)$ time evolution of the particle current. For $(a)$ to $(e)$, the value at $t=0$ is subtracted.
The evolution parameters are the same as in Fig.~\ref{fig:comp_vs_time}. 
\label{fig:contribution_current}}
\end{figure} 

To understand better the complex dynamics at play in the presence of regions close to filling one, 
we show in Fig.~\ref{fig:comp_vs_time} real-time snapshots of the compressibility and particle current 
for the ramp time $\tau=7\hbar/J$. 
The connection between these two quantities becomes evident at close inspection.
One first notices that, once again, the compressibility in regions away from unit filling evolves 
quickly while the flow of atoms towards the system boundaries takes a much longer time to set in.
In addition, once the compressibility is sufficiently suppressed in the regions of filling one, the 
current in these regions weakens, which slows down the density redistribution across the gas. 
Even though the regions close to unit filling are small and are not real Mott-insulating plateaus, they 
still reduce significantly the transport from the inner to the outer superfluid domains. 
Consequently, the onset of low compressibility regions explains why systems above unity filling evolve 
slowly when $U$ is increased.
From here on, we will refer to these regions as ``Mott barriers''.

To shed even more light on the build up and suppression of the particle current, we analyze the contribution of the different 
terms appearing in \eqref{eq:diffcurrent} which make up the time-derivative of the particle current. 
For each term, we plot in Fig.~\ref{fig:contribution_current} 
the different contributions to $-\text{div}\moy{\hat{J}^j}$ for various times 
in order to understand what drives the evolution of the particle current.
The first remarkable feature is Fig.~\ref{fig:contribution_current}(e) where the 
time-derivative of the current changes sign near unity filling around $t\simeq 5\hbar/J$. 
This inversion is a clear indication that the current is being suppressed by the formation of Mott barriers.
By considering each contribution separately using Fig.~\ref{fig:contribution_current}(a-d), 
we observe that the main driving terms boosting the particle currents are the ones related to the 
local kinetic energy \eqref{eq:contr_mu} and \eqref{eq:contr_corr2}, and the density assisted hoppings 
term \eqref{eq:contr_densassisted} while the density gradients \eqref{eq:contr_n} become significant only 
at the edges where the density varies rapidly.
The most striking phenomenon is due to the density assisted hoppings term \eqref{eq:contr_densassisted}. 
We see on Fig.~\ref{fig:contribution_current}(d) that this term, which is non-zero only 
in the presence of interactions, changes sign in the regions where Mott barriers are 
forming thus drastically slowing down the equilibrating out-flow of atoms. 

We finally stress again that the time-scales associated with the 
contributions~\eqref{eq:diffcurrent} are essentially controlled by the steepness of profiles induced by the trapping potential $V_0$. 
Within our choice of parameters they take longer times than the intrinsic evolution.
Experimentally, changing the confinement strength $V_0/J$ would affect both in time and magnitude 
the creation of Mott barriers~\cite{Note3}.

\subsubsection{Energy transport and heat production}
\begin{figure}[t]
\centering
\includegraphics[width=\columnwidth,clip=true]{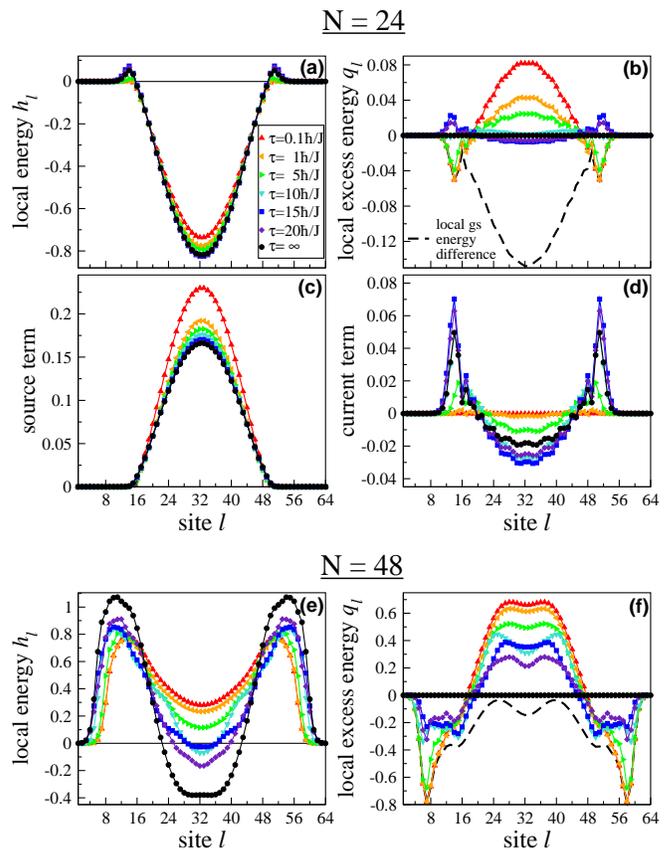}
 \caption{(color online). Final local energy $(a,e)$ and local excess energy $(b,f)$ profiles after a 
slow quench from $U_i=4J$ to $U_f=6J$ for different ramp times, $\tau$, and for the ground state ($\tau=\infty$)
at $U=6J$ for a trapped system. $(c)$ is the contribution 
to $q_l$ due to the external operator
(see \eqref{eq:heatfluct}) while $(d)$ is the contribution from the energy currents (see \eqref{eq:heatdiv}).
The dashed line in $(b)$ and $(f)$ corresponds to $\moy{\hat{h}_l}_{0,i} - \moy{\hat{h}_l}_{0,f}$. 
\label{fig:trap_profiles_locale_4_6}}
\end{figure}

We now turn our attention to the energy transport and heat production during a quench.
We present in Fig.~\ref{fig:trap_profiles_locale_4_6}(a, e) the final local energy profiles
for $N=24$ and $48$~\cite{Note4}.
This figure confirms our findings obtained from the analysis of the local density 
and compressibility profiles: for $N=24$ the system approaches the adiabatic limit much faster than for $N=48$. 
For $N=48$, the final profile remains highly excited even for the longest ramp time considered ($\tau \approx 25 \hbar/J$).
The local excess energy production highlights a series of differences between the evolution of systems with 
filling below and above one (see Fig.~\ref{fig:trap_profiles_locale_4_6}(b) and Fig.~\ref{fig:trap_profiles_locale_4_6}(f)). 
We first notice that the local excess energy is smaller by nearly an order of magnitude for $N=24$ compared to $N=48$.
Furthermore, while for $\tau=25\hbar/J$ and $N=24$ the local excess energy is rather uniformly distributed 
and close to zero, the $N=48$ result exhibits strong spatial fluctuations with $q_l$ large and negative at 
the edges and large and positive at the center of the cloud. In fact, the local excess energy pattern 
resulting from the quench is highly non-trivial even for the seemingly simplest situation where $N=24$, as 
illustrated in Fig.~\ref{fig:trap_profiles_locale_4_6}(b, c, d).
At short ramp times (sudden quench limit), particles and energy currents are negligible so that the 
term \eqref{eq:heatdiv} does not contribute, all the final excess energy being a balance between the ground 
state energy difference \eqref{eq:heatdiff} and the density fluctuations average \eqref{eq:heatfluct}.
The latter is always positive and distributed rather uniformly in a Gaussian-like function whose maximum 
decreases with $\tau$ (see Fig.~\ref{fig:trap_profiles_locale_4_6}(c)). 
Hence, for short ramp times, the bulk retains most of the local excess energy while the 
edges have negative $q_l$ due to the term \eqref{eq:heatdiff}.
For longer ramp times, energy currents set in with the effect of redistributing energy from the bulk to the 
edges (see Fig.~\ref{fig:trap_profiles_locale_4_6}(b)).
Thus, these currents tend to strongly reduce both the spatial fluctuations and the total excess energy (heat) produced 
by the quench. For intermediate times, either negative or positive $q_l$ at the edges and 
in the bulk (see for instance the opposite distribution for $\tau = 10\hbar/J$ and $\tau = 15\hbar/J$ for $N=24$) can be found. 
This effect arises as, for these parameters, the density profiles overshoot their final ground state configurations. 
For $N=48$, the ``Mott barriers effect'' tends to freeze 
the excess local energy pattern to the sudden quench typical distribution with negative $q_l$ at the edges 
and positive in the bulk. Let us note that the freezing of the local excess energy pattern is strongly 
related to the frozen density pattern. 

\begin{figure}[t]
\centering
\includegraphics[width=\columnwidth,clip=true]{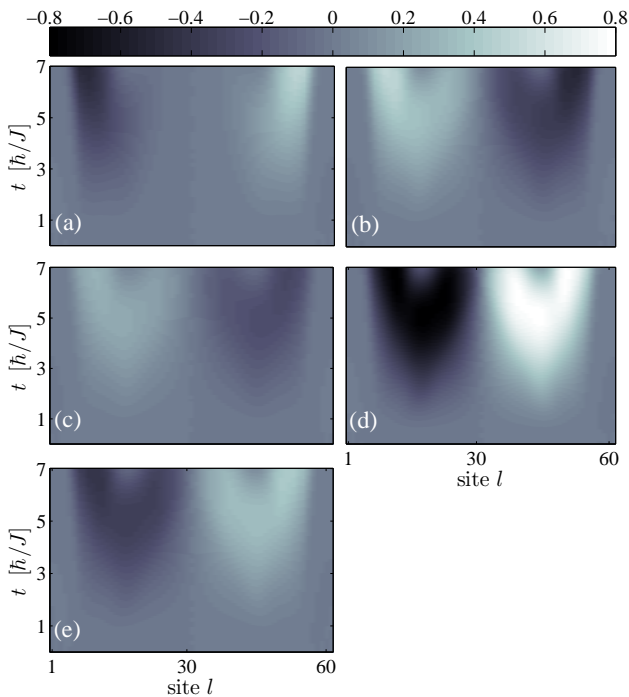}
  \caption{Time-evolution of the different contributions to the energy current:
  $(a)$ Eq.~\eqref{eq:energy_mu}, $(b)$ Eq.~\eqref{eq:energy_j}, $(c)$ 
  Eq.~\eqref{eq:energy_corr2} and
  $(d)$ Eq.~\eqref{eq:energy_densass}. Time evolution of the full energy current $(e)$.
  The evolution parameters are the same as in Fig.~\ref{fig:comp_vs_time}. 
  \label{fig:contr_energy}}
\end{figure}

Looking at the contributions to the evolution of the energy current in Fig.~\ref{fig:contr_energy}, 
we can identify the leading contribution driving the energy redistribution. 
Comparing Fig.~\ref{fig:contribution_current}(f) and \ref{fig:contr_energy}(e), we see that the overall 
evolution of the particle and energy current is very similar: 
both currents set in at about the same time, and are suppressed in regions where Mott barriers form. 
We also observe that the main contribution, determining the sign, to the energy current is from the density-assisted 
particle current (see Fig.~\ref{fig:contr_energy}(d)).
However, this flow of energy towards the system edges is partially counterbalanced by two terms (Fig.~\ref{fig:contr_energy}(b) 
and (c)) where the energy transport
occurs in the opposite direction to the particle current. 

\begin{figure}[t]
\centering
\includegraphics[width=0.9\linewidth,clip=true]{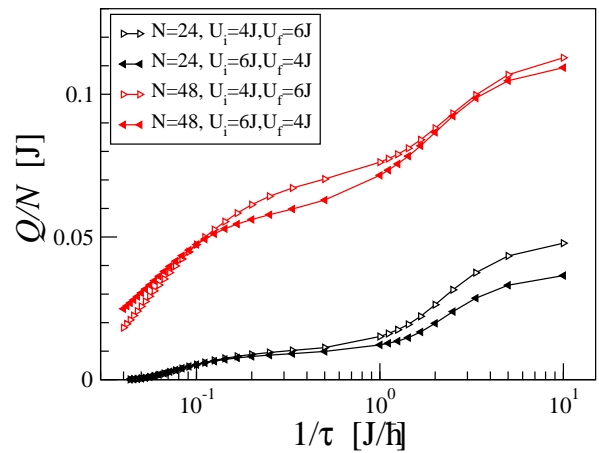}
\caption{(color online). Total heat $Q(\tau)$ vs. inverse ramp time $\tau$ for both directions of the quenches for $N=24$ and $N=48$.
\label{fig:totheat}}
\end{figure}

Finally, one may wonder how the total heat produced $Q$ as a function of $1/\tau$ differs from the 
homogeneous situation studied in Ref.~\onlinecite{BernierKollath2011}.
We show how the heat behaves as a function of the ramp time in Fig.~\ref{fig:totheat} for $N=24$ and $48$ for both the 
quench from $U_i=4J$ to $U_f=6J$ and its reverse.
We first notice that 
these curves are more complex than the one presented in Ref.~\onlinecite{BernierKollath2011} for a homogeneous system.
We also observe that the heat per atom produced in the case $N=24$ is always much lower than the one 
for $N=48$. Our understanding of this phenomenon is that for lower filling the populated excited 
states are less energetic as they are less likely to have doubly occupied sites. 
We finally observe that, for fast ramps, the heat produced in the protocol with $U_i=4J$ and $U_f=6J$ is larger 
than for the reverse protocol (while the opposite happens for slower ramps). We relate this to the fact that in the $U_i/J=4$ 
initial state a lot of particle fluctuations are present leading to a large interaction energy in the final state.
However, in order to fully understand the crossover to the inverse behavior at slow ramp times, the number of excitations 
that are created and their final energies
would need to be identified, a task that we leave to future studies.
 
\subsubsection{Comparison with mean-field Gutzwiller method}
\label{sec:gutz}

\begin{figure}[t]
\centering
\includegraphics[width=0.9\columnwidth,clip=true]{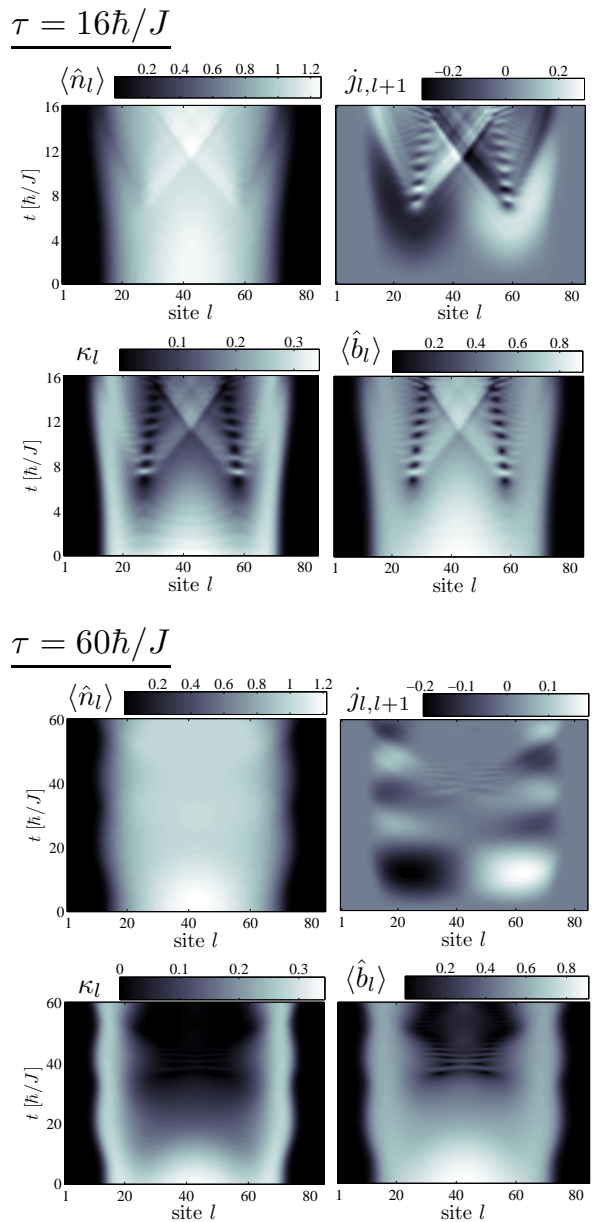}
  \caption{Time-evolution, within the Gutzwiller method, of a trapped one-dimensional Bose gas loaded 
           into an optical lattice during a quench from $U_i=6J$ to $U_f=15J$ for two different ramp times. Plotted
           quantities are the local density $\langle \hat{n}_l \rangle$, the local compressibility $\kappa_l$, 
           the current $j_{l,l+1}$ and the superfluid order parameter $\langle\hat{b}_l \rangle$. 
           $N=54$, $L=84$, $V_t=0.006J$. Upper panels: $\tau=16\hbar/J$. Lower panels: $\tau=60\hbar/J$. 
           \label{fig:gutz}}
\end{figure}

Our aim here is to understand to what extent the mean-field Gutzwiller method can 
describe the time-evolution of
a Bose gas loaded into a one-dimensional optical lattice and confined to a parabolic trap. With this objective
in mind, we study here a system made of $54$ atoms confined to a parabolic trap with $V_0=0.006J$ and loaded in an 
optical lattice of $84$ sites, and consider slow quenches from $U_i = 6J$ to $U_f = 15J$ for two different 
ramp times: $\tau=60\hbar/J$ and $\tau=16\hbar/J$.
These quenches begin on the superfluid side and the interaction strength is linearly increased up to a value 
above the $n=1$ homogeneous superfluid-Mott-insulating transition, occurring at $U_c \approx 11.7J$ (using the Gutzwiller method). 
At mean-field level, the ground state at $U = 6J$ is a superfluid with a 
central density above one while the ground state at $U = 15J$ presents a broad Mott plateau. 

Considering Fig.~\ref{fig:gutz}, we first notice that the Gutzwiller method captures well the presence of 
two dynamical regimes. For both ramps, we see that the evolution of the local compressibility and superfluid order
parameter begins at $t = 0$ whereas the local density and the particle current remain fixed to their initial 
values for a few $\hbar/J$. For a sufficiently fast quench, as shown in the upper panels of Fig.~\ref{fig:gutz},
we see that the superfluid order parameter, the compressibility and the current are strongly suppressed in a narrow
region around filling one. The local suppression of these three quantities around $ t = 6 \hbar/J$ signals the formation
of Mott barriers hindering the flow of atoms. 
For $\tau=16J/\hbar$, these barriers are unstable and we notice the presence of oscillations reminiscent of the
ones arising when a strongly interacting phase is abruptly quenched to strong interactions \cite{Greiner2002b}. 

By comparison, for sufficiently slow ramps, a stable Mott-insulating plateau forms at long times. On this plateau, the 
condensate order parameter and the compressibility drop to zero. This total suppression of the density 
fluctuations is an artifact of the mean-field method and also results in the absence of particle current on the plateau
as, within the Gutzwiller picture, the current factorizes into $j_{l-1,l}= 2J~\Im (\aver{\bhat_{l-1}}\!^* \aver{\bhat_l})$.
Finally, we also observe in the lower panels of Fig.~\ref{fig:gutz} that the quench triggers collective 
breathing modes signaled by density oscillations
(along the time axis) in boundary regions~\cite{Note5}.

From this discussion of slow superfluid-Mott-insulating quenches within the Gutzwiller method, we conclude that this
approach captures some of the important out-of-equilibrium physical phenomena uncovered by t-DMRG, however as expected it cannot
provide an accurate quantitative picture. 

\subsection{Evolution of non-local quantities}
\label{sec:corr}

\subsubsection{Real-space correlations}

\begin{figure}[t]
\centering
  \includegraphics[width=\columnwidth,clip=true]{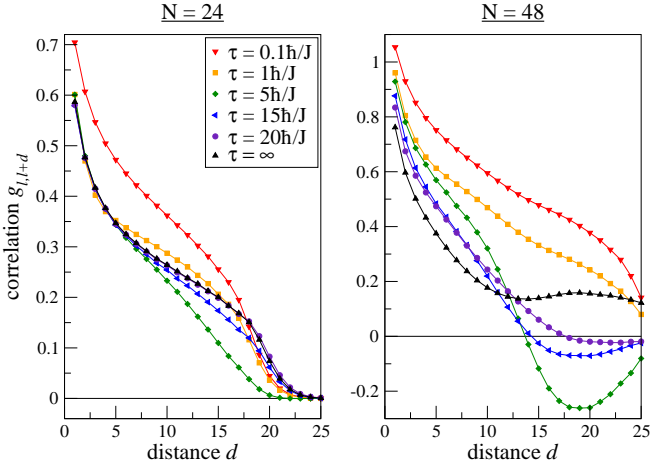} 
  \caption{(color online). Value of the correlator $g_{l,l+d}$ with $l=L/2+1$ after a 
           slow quench from $U_i = 4J$ to $U_f = 6J$ in a trap for
           various ramp times $\tau$ and the final ground state ($\tau=\infty$). 
           Left panel: $N=24$. Right panel: $N=48$.
           \label{fig:trap_corr_long_range_MI}}
\end{figure}

Local and non-local correlations can propagate very differently during a quench. 
To understand how correlations evolve during the slow quench of a global parameter, 
we investigate here the evolution of single particle correlations. 
Past studies on other systems have found that, after a slow parameter change, long-range
correlations take a long time to adjust~\cite{RodriguezSantos2009, ClarkJaksch2004}. 
For example, for spin systems described by locally acting Hamiltonians, the propagation
of correlations during the slow change of a global parameter was found to be bound by a 
``light-cone''. Outside of this light-cone, the so-called Lieb-Robinson bound, only 
exponentially small changes to the correlations can be detected~\cite{LiebRobinson1972,BravyiVerstraete2006}. 

We show in Fig.~\ref{fig:trap_corr_long_range_MI} 
the value of the correlator $g_{L/2+1,L/2+1+d}$ at $t = \tau$ for different ramp times and two fillings. 
The first striking result emerging from our study is that the evolution of this correlator is not
monotonic with the ramp time. We also find that in all cases the short distance correlator
responds quickly to the increase of the interaction strength, and that even for the fastest quenches
the final correlation function differs considerably from the initial ground state correlator.
Focusing on the left panel of Fig.~\ref{fig:trap_corr_long_range_MI}, we see that at low filling ($N=24$) 
the short distance correlations reach their final ground state
values for almost all considered ramp times. In contrast, the longer range correlations 
take much longer to reach their corresponding ground state values.
For example, for $\tau= 5 \hbar/J$, the long distance correlations have clearly not yet relaxed to their
final ground state values.

In the situation where regions with filling above one are present (right 
panel of Fig.~\ref{fig:trap_corr_long_range_MI}), the evolution is even more involved. In this 
case, the correlator at $t = \tau$ varies non-monotonically with distance and takes negative values
for intermediate ramp velocities. Even for the slowest ramps considered, the correlator
deviates considerably from its final ground state value at all distances. Finally, let us note that the
final ground state correlations present a dip at a distance corresponding to the location of 
regions of filling one.

\subsubsection{Interference pattern for experiments}
\label{sec:interf}

\begin{figure}[t]
\centering
  \includegraphics[width=\columnwidth,clip=true]{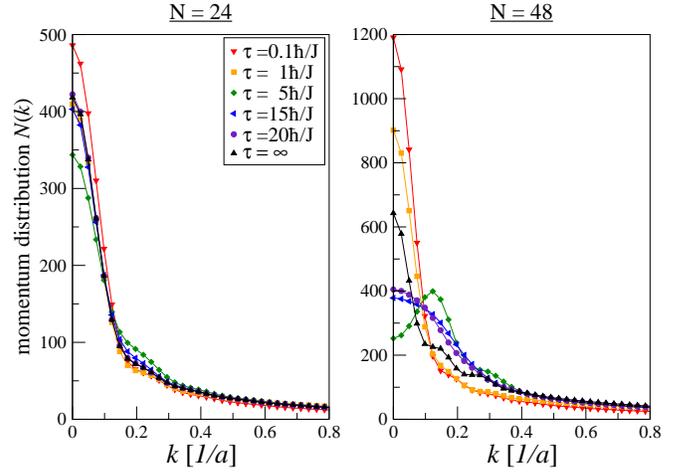} 
 \caption{(color online). Final value of the interference pattern $N(k)$ after a slow quench 
          from $U_i = 4J$ to $U_f = 6J$ in a trap for different ramp times 
           $\tau$ and for the final ground state ($\tau=\infty$).  
           Left panel: $N=24$. Right panel: $N=48$.
 \label{fig:trap_interf}}
\end{figure}

Part of the complex dynamical behavior presented above can be observed experimentally in the 
time-of-flight interference pattern $N(k)$ defined in \eqref{eq:nk}.
As shown on the left panel of Fig.~\ref{fig:trap_interf}, at low filling, the interference pattern present a peak at
$k = 0$ which changes in amplitude non-monotonically with the ramp time. This behavior 
reflects the non-monotonic variation of long range correlations discussed in the previous section. 
The final interference pattern is very different in the presence of regions close to filling one. In this case,
a peak at $k \neq 0$ develops at intermediate ramp times (see the right panel of Fig.~\ref{fig:trap_interf}). 
This peak signals the strong out-of-equilibrium character of the state formed during the slow quench. However,
the absence of such a peak cannot be used to conclude that the system evolves adiabatically. Unfortunately, the
interference pattern is not as sensitive to out-of-equilibrium features as correlation functions are: $N(k)$ 
can be dominated by large ``in-equilibrium'' contributions coming from the short range correlators.

\section{Melting of Mott-insulating regions}
\label{sec:MI_SF}

In this section, we consider a linear quench from $U_i=6J$ to $U_f=4J$. At $U_i$, the system presents a sizable 
Mott-insulating ``shell'' and a superfluid center, while $U_f$ is close to the homogeneous superfluid-Mott-insulating transition 
point. The ground state density and compressibility profiles at $U_f$ show none of the features associated 
with the presence of Mott regions. Here again we focus on the different aspects of the dynamics:
time-scales, particle transport, energy production, and experimental signatures.

\subsection{Existence of two dynamical regimes}

\begin{figure}[t]
\centering
\includegraphics[width=\columnwidth,clip]{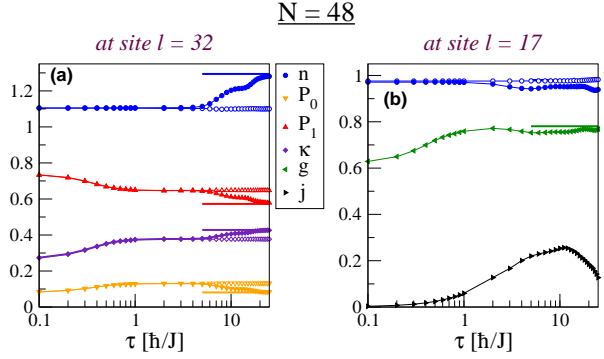}
\caption{(color online). Slow quench from $U_i = 6J$ to $U_f = 4J$, $N=48$.
           Evolution of local observables in the presence of a trap as a function of the ramp time $\tau$, 
           and compared with that of a homogeneous system (open symbols) having the same initial local density. 
           Observables are the density $n_l$, compressibility $\kappa_l$, occupancy probabilities $P_0$ and $P_1$, 
           neighboring correlation $g_{l,l+1}$ and particle current $j_{l,l+1} = \moy{\hat{j}_{l,l+1}}$.
  \label{fig:n_vs_ts_melt}}
\end{figure}

When the interaction strength is lowered, the dynamics at play are also characterized by ``two dynamical regimes''. 
However, as seen on Fig.~\ref{fig:n_vs_ts_melt}, in this case, the atoms are moving towards the center of
the system not towards the edges.

\subsection{Density, compressibility and energy profiles}

\begin{figure}[t]
\centering
\includegraphics[width=\columnwidth,clip=true]{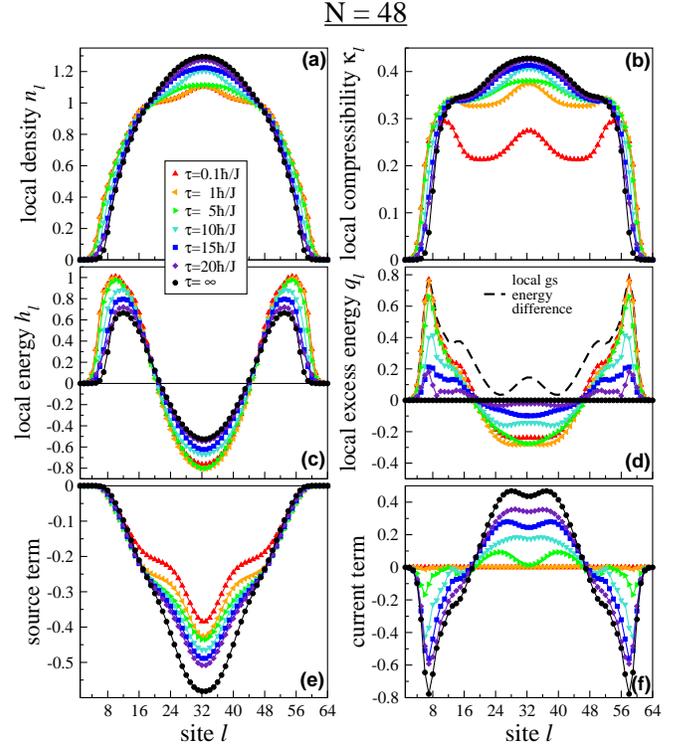}
 \caption{(color online). Final profiles for the local density $(a)$, local compressibility $(b)$, local energy $(c)$ and 
          local excess energy $(d)$ after a slow quench from $U_i=6J$ to $U_f=4J$ for different ramp 
          times, $\tau$, and for the ground state ($\tau=\infty$)
          at $U=4J$ for a trapped system. $(e)$ is the contribution to $q_l$ due to the external operator
          (see \eqref{eq:heatfluct}) while $(f)$ is the contribution from the energy currents (see \eqref{eq:heatdiv}).
          The dashed line in $(d)$ corresponds to $\moy{\hat{h}_l}_{0,i} - \moy{\hat{h}_l}_{0,f}$.
\label{fig:trap_profiles_6_4}}
\end{figure}

Considering  the density and compressibility profiles for different ramp times shown in Fig.~\ref{fig:trap_profiles_6_4},
we notice that for ramp times of the order of $5\hbar/J$ the Mott-insulating regions are 
almost fully melted and that the system is more
compressible. For example, local dips, initially present, have completely disappeared and only plateaus remain. For even 
longer ramp times, the final density and compressibility profiles resemble closely the
$U_f$ ground state. Density redistribution occurs at a much faster pace when Mott-insulating regions are melted away than when they
are formed since in the former case Mott barriers are no longer effective. To conclude this comparison 
between the two protocols,
it is interesting to note that, when the interaction strength is lowered, energy is transferred from the edges to the
center of the system as atoms pile up in the central region. The opposite occurs when the interaction is 
increased.

\begin{figure}[t]
\centering
\includegraphics[width=\columnwidth,clip=true]{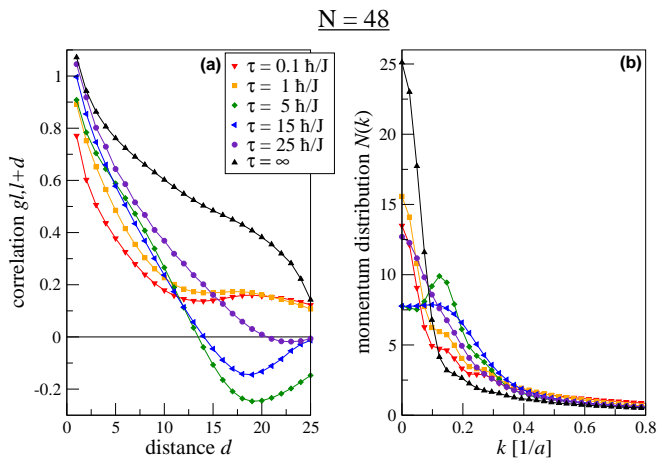}
\caption{(color online). Ramp from $U_i=6J$ to $U_f=4J$, $N=48$. Left panel: final value 
  for correlator $g_{l,l+d}$ with $l=L/2+1$ for different ramp times, $\tau$, and the
  final ground state ($\tau=\infty$). Right panel: final value for the interference pattern $N(k)$ (see \eqref{eq:nk}) for
  different ramp times and the final ground state ($\tau=\infty$).
  \label{fig:trap_corr_long_range_6_4}}
\end{figure}

\subsection{Real space correlations and interference patterns}

Even though for long ramp times the density and compressibility profiles seem to evolve almost adiabatically, 
the single particle correlator $g_{L/2+1,L/2+1+d}$ indicates that the system
is still far from equilibrium. As we can see on Fig.~\ref{fig:trap_corr_long_range_6_4}(a), this correlator is
negative at large distances and remains far from its ground state $U_f$ value even at long ramp times,
except for short distances. Therefore, to judge if the system has reached equilibrium by solely considering the
density and compressibility profiles is inadequate. Our results show unequivocally that the system is far from
having fully relaxed even at long ramp times. The non-equilibrium nature of the final state can be partially 
probed by measuring experimentally the interference pattern (see~\eqref{eq:nk}). 
On Fig.~\ref{fig:trap_corr_long_range_6_4} (right), we see that for intermediate ramp times, 
the peak at $k = 0$
is shifted to higher momentum signaling the
non-equilibrium nature of the final state. However, as the interference pattern is a sum over all-distance 
correlators and is dominated by short-distance values, it is difficult to distinguish the shifted peak 
at long ramp times.

\section{Conclusion}

In this article we investigated the dynamics of the Mott-insulating regions of a bosonic gas
trapped and loaded into an optical lattice as the interaction strength is changed linearly with time.
We considered two situations: we first studied how Mott domains are formed by ramping up the 
interaction strength from $U_i=4J$ to $U_f=6J$ and, in a separate set of simulations, 
investigated how the domains melt when $U$ is ramped down. We conducted this study by examining
how the atomic density and compressibility profiles evolve, how particles and energy flow through the
system, how heat is produced and how single particle correlations propagate as a function of the ramp
time. For both situations we confirmed the existence of two dynamical regimes: an intrinsic regime
occurring at short times before particle transport sets in, and a long time behavior connected to
the system inhomogeneities and controlled by the strength of the underlying trapping potential.
We were able to establish the existence of these regimes on firmer grounds using various arguments
based on time-dependent perturbation theory. In a system with regions above unity filling, we found that a linear
increase of the interaction strength is accompanied by the formation of Mott insulating barriers
which hinder the flow of atoms from the center towards the edges. The emergence of these barriers is
evidenced by dips in the local compressibility and by the suppression of the particle current
in regions where the local density nears unity. We also established that, in these regions, 
the change in sign of the particle current time-derivative is due to density assisted hopping,
a mechanism which only exists when $U$ is finite. The presence of theses barriers has multiple
consequences, among others, the system ``freezes'' into a highly excited configuration and 
long range single particle correlations deviate strongly from their final ground state values, 
even for the slowest ramp considered. This last feature could possibly be detected 
experimentally from the gas interference pattern. By comparison, when the interaction strength is ramped down
the evolution is much less involved. For sufficiently long ramps, the density, compressibility and
local energy profiles approach the corresponding $U_f$ ground state configuration. However, even for the
slowest ramp considered, the final system is still far from being equilibrated as the associated one-body 
correlator departs strongly from its final ground state value for all distances.
To conclude, we believe that this thorough investigation of the dynamics of a strongly interacting 
bosonic gas will help experimentalists devise protocols to prepare complex quantum phases and provides 
a new perspective to the understanding of the coherent evolution of quantum systems in the presence of
inhomogeneities.

\acknowledgments 
We are grateful to R.~Citro, S.~Natu, E.~Orignac and A.~Rosch for fruitful discussions. 
We acknowledge financial support from the Triangle de la Physique,
the Agence Nationale de la Recherche (under contract FAMOUS), the 
SNSF (under division II), the DARPA-OLE program, and the Canadian
Institute for Advanced Research. Financial support for the computer 
cluster on which some of the calculations were performed has been provided 
by the Fondation Ernst et Lucie Schmidheiny.

\appendix
\section{Useful commutators}
\label{app:commutators}

We list below some helpful commutators which generate the terms appearing in the various 
current operators of Sec.~\ref{sec:continuity}:
\begin{align*}
\com{\bhat_l}{\hat{n}_l}                 &= \bhat_l                     \\
\com{\bhat^{\dag}_k \bhat_l}{\hat{n}_l}      &= \bhat^{\dag}_k \bhat_l         \\
\com{\bhat^{\dag}_k \bhat_l}{\hat{n}_l^2}    &= (1+2\hat{n}_l)\bhat^{\dag}_k \bhat_l \\ 
\com{\bhat^{\dag}_l \bhat_k}{\bhat^{\dag}_m \bhat_l} &=-\bhat^{\dag}_m \bhat_k \quad(m\neq k)\\
\com{\bhat^{\dag}_l \bhat_k}{\bhat^{\dag}_k \bhat_l} &= \hat{n}_l-\hat{n}_k
\end{align*}
Permutations between $l$ and $k$ are obtained by taking the hermitian conjugate.

\section{Equation of evolution for the occupancy probability}
\label{app:proba}

A single-site $l$ of the Bose-Hubbard model is fully characterized by the 
occupancy probabilities $P_{n_l}$ of having $n_l$ bosons onsite. 
The reduced density-matrix of the site is diagonal because of the conservation of the total number of bosons and reads
\begin{eqnarray}
\rho_{l}(t) &=& \sum_{n_l=0}^N P_{n_l}(t)\ket{n_l}\bra{n_l}\,.
\end{eqnarray}
The mean-value and standard deviation of the $P_{n_l}(t)$ distribution are simply $\moy{n_l}(t)$ and $\sqrt{\moy{\kappa_l}(t)}$.
In order to get the continuity equation for $P_{n_l}(t)$, we introduce the characteristic function 
\begin{equation}
\label{eq:generating-function}
f(\theta;t) = \moy{e^{i\theta \hat{n}_l}}(t) = \sum_{p=0}^{+\infty}\frac{(i\theta)^p}{p!}\moy{\hat{n}^p_l}(t) 
						= \sum_{n_l} P_{n_l}(t)e^{i\theta n_l}
\end{equation}
such that $\moy{\hat{n}^p_l}(t) = (-i)^p\left.\frac{d^p f}{d\theta^p}\right\vert_{\theta=0}$. Using \eqref{eq:generating-function}, 
the probabilities are recovered using 
\begin{equation}
P_{n_l}(t) = \frac{1}{2\pi}\int_{0}^{2\pi}\!d\theta \;e^{-i\theta n_l}f(\theta;t).
\label{eq:PnFourier}
\end{equation}
Using the relations $\bhat_le^{z\hat{n}_l} = e^{z(\hat{n}_l+1)}\bhat_l$ and $\bhat^{\dag}_le^{z\hat{n}_l} = e^{z(\hat{n}_l-1)}\bhat_l^{\dag}$, we get
\begin{widetext}
\begin{eqnarray}
\hbar\partial_t \moy{e^{i\theta \hat{n}_l}} &=& iJ(e^{i\theta}-1)\left\{
\moy{\bhat^{\dag}_{l}\bhat_{l-1}e^{i\theta \hat{n}_l}} - \moy{e^{i\theta \hat{n}_l}\bhat^{\dag}_{l-1}\bhat_{l}}
-[\moy{e^{i\theta \hat{n}_l}\bhat^{\dag}_{l+1}\bhat_{l}} - \moy{\bhat^{\dag}_{l}\bhat_{l+1}e^{i\theta \hat{n}_l}}]
\right\}\\
&=& iJ\left\{(e^{i\theta}-1)[\moy{\bhat^{\dag}_{l}\bhat_{l-1}e^{i\theta \hat{n}_l}} +\moy{\bhat^{\dag}_{l}\bhat_{l+1}e^{i\theta \hat{n}_l}}]
 - (1-e^{-i\theta}) [\moy{\bhat^{\dag}_{l-1}\bhat_{l}e^{i\theta \hat{n}_l}}+\moy{\bhat^{\dag}_{l+1}\bhat_{l}e^{i\theta \hat{n}_l}} ]
\right\}
\label{eq:generating-function-evol}
\end{eqnarray}
\end{widetext}
By taking the $p^{\text{th}}$ derivatives of this equation with respect to $\theta$, we get the time-evolution 
of $\moy{n_l^p}$. In particular, we recover the time-evolution of the local density with the first derivative. 
We also see that this equation yields correlated currents of the form $\moy{\bhat^{\dag}_{l}\bhat_{l+1} n_l^{p-1}}$ 
for the evolution of $\moy{n_l^p}$.
If we want the time-evolution of the probabilities $P_{n_l}(t)$, we have to integrate \eqref{eq:generating-function-evol} 
over $\theta$. Formally, we have:
\begin{equation}
\partial_t P_{n_l} = \frac{1}{2\pi}\int_{0}^{2\pi}d\theta\, e^{-i\theta n_l} \partial_t \moy{e^{i\theta \hat{n}_l}}\;.
\label{eq:PnEvolution}
\end{equation}
While these formulas are of little help for the numerics, they highlight the connection between the evolution of 
the $P_{n_l}$ distribution and the correlated currents. 

Furthermore, knowing the evolution equation of $P_{n_l}(t)$ allows one to obtain the evolution equation for 
the associated onsite entropy of particle fluctuations 
\begin{equation}
s_l(t) = -k_B \sum_{n_l} P_{n_l}(t) \ln P_{n_l}(t)
\label{eq:local-entropy}
\end{equation}
This entropy is rigorously defined also in the non-equilibrium regime. 
Indeed, in a superfluid regime, or even for free bosons where $P_{n_l}$ is Poissonian, many $n$ have significant 
weights leading to large $s_l$, while the $n=1$ Mott regime is such that there is a strong peak at $n=1$ with 
shoulders at $n=0,2$, associated with a much smaller entropy. Thus, $s_l$ is sensitive to the nature of the 
local phase/domain.
Formally, we immediately get the equation of evolution from
\begin{equation}
\partial_t s_l = -k_B\sum_{n_l} (\partial_t P_{n_l})\ln P_{n_l}\;.
\end{equation}

\end{document}